\documentclass{article}
\usepackage{latexsym}
\usepackage{psfig}
\newcommand{\half}{\mbox{$\textstyle \frac{1}{2}$}}
\newcommand{\ket}[1]{\left | \, #1 \right \rangle}
\newcommand{\bra}[1]{\left \langle #1 \, \right |}

\begin{document}
\rightline{\vbox{\baselineskip=12pt{\hbox{Contribution to Complexity}}}}
\bigskip
\centerline{\Large \bf On quantum algorithms}
\bigskip\bigskip
\centerline{\large Richard Cleve, Artur Ekert, Leah Henderson,
Chiara Macchiavello and Michele Mosca}
\medskip
\centerline{\it Centre for Quantum Computation}
\centerline{\it Clarendon Laboratory, University of Oxford, Oxford OX1
3PU, U.K.}
\medskip
\centerline{\it Department of Computer Science}
\centerline{\it University of Calgary, Calgary, Alberta, Canada T2N 1N4}
\medskip
\centerline{\it Theoretical Quantum Optics Group}
\centerline{\it Dipartimento di Fisica ``A. Volta'' and I.N.F.M. -
  Unit\`a di Pavia}
\centerline{\it Via Bassi 6, I-27100 Pavia, Italy}
\bigskip
%\centerline{\large \bf Version 2.0}
%\centerline{Tuesday, 14 October 1997}
%\medskip

%---------------------------------------------

\begin{abstract}
Quantum computers use the quantum interference of different computational
paths to enhance correct outcomes and suppress erroneous outcomes of
computations.
In effect, they follow the same logical paradigm as (multi-particle)
interferometers.
We show how most known quantum algorithms, including quantum algorithms
for factorising and counting, may be cast in this manner.
Quantum searching is described as inducing a desired relative phase between
two eigenvectors to yield constructive interference on the sought elements
and destructive interference on the remaining terms.
\end{abstract}

\parskip=5pt %plus 2pt minus 1pt
%---------------------------------------------

\section{From Interferometers to Computers}

Richard Feynman~\cite{Feynman} in his talk during the First Conference
on the Physics of Computation held at MIT in 1981 observed that it
appears to be impossible to simulate a general quantum evolution on a
classical probabilistic computer in an {\em efficient} way.  He
pointed out that any classical simulation of quantum evolution appears
to involve an exponential slowdown in time as compared to the natural
evolution since the amount of information required to describe the
evolving quantum state in classical terms generally grows
exponentially in time.  However, instead of viewing this as an
obstacle, Feynman regarded it as an opportunity.  If it requires so
much computation to work out what will happen in a complicated
multiparticle interference experiment then, he argued, the very act of
setting up such an experiment and measuring the outcome is tantamount
to performing a complex computation.  Indeed, all quantum
multiparticle interferometers {\em are} quantum computers and some
interesting computational problems can be based on estimating internal
phase shifts in these interferometers.  This approach leads to a
unified picture of quantum algorithms and has been recently discussed
in detail by Cleve {\it et al.}~\cite{CEMM}.

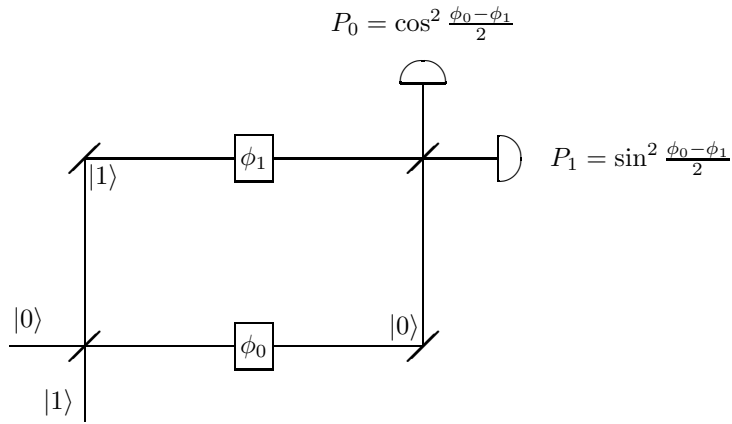
\begin{figure}
\centering
\setlength{\unitlength}{1mm}
\begin{picture}(80,70)
%general outline
\put(0,30){\line(1,0){30}}
\put(30,27){\framebox(5,6){$\phi_0$}}
\put(35,30){\line(1,0){20}}
\put(10,20){\line(0,1){35}}
\put(55,30){\line(0,1){35}}
\put(10,55){\line(1,0){20}}
\put(30,52){\framebox(5,6){$\phi_1$}}
\put(35,55){\line(1,0){30}}
%photodetectors
\put(65,52){\line(0,1){6}}
\put(65,55){\oval(6,6)[r]}
\put(52,65){\line(1,0){6}}
\put(55,65){\oval(6,6)[t]}
%probability labels
%\put(75,67){\vector(-1,0){15}}
\put(40,70){\makebox(30,6){$P_0=\cos^2\frac{\phi_0-\phi_1}{2}$}}
%\put(75,55){\vector(-1,0){5}}
\put(69,52){\makebox(30,6){$P_1=\sin^2\frac{\phi_0-\phi_1}{2}$}}
%beam-splitters
{\thicklines
\put(8,28){\line(1,1){4}}
\put(53,53){\line(1,1){4}}
%mirrors
\put(8,53){\line(1,1){4}}
\put(53,28){\line(1,1){4}}}
%input labels
\put(0,31){\makebox(5,5){$|0\rangle$}}
\put(4,20){\makebox(5,5){$|1\rangle$}}
%intermediate labels
\put(50,30){\makebox(5,5){$|0\rangle$}}
\put(10,50){\makebox(5,5){$|1\rangle$}}
\end{picture}
\label{fig-MZ}
  \caption{A Mach-Zehnder interferometer with two phase shifters.  The
  interference pattern depends on the difference between the phase
  shifts in different arms of the interferometer.}
\end{figure}

Let us start with the textbook example of quantum interference, namely
the double-slit experiment, which, in a more modern version, can be
rephrased in terms of Mach-Zehnder interferometry (see
Fig.~\ref{fig-MZ}).

A particle, say a photon, impinges on a beam-splitter (BS1), and, with
some probability amplitudes, propagates via two different paths to
another beam-splitter (BS2) which directs the particle to one of the
two detectors.  Along each path between the two beam-splitters, is a
phase shifter (PS).  If the lower path is labelled as state $\ket{0}$
and the upper one as state $\ket{1}$ then the particle, initially in
path $\ket{0}$, undergoes the following sequence of transformations
\begin{eqnarray}
\ket{0}&\stackrel{\mbox{\tiny BS1}}{\longrightarrow}& \frac{1}{\sqrt 2}
\left(\ket{0} + \ket{1}\right) \nonumber \\
&\stackrel{\mbox{\tiny
PS}}{\longrightarrow}& \frac{1}{\sqrt 2} (e^{i\phi_{0}}\ket{0}
+e^{i\phi_{1}}\ket{1}) = e^{i\frac{\phi_{0}+\phi_{1}}{2}} \frac{1}{\sqrt
2} (e^{i\frac{\phi_{0}-\phi_{1}}{2}}\ket{0} +
e^{-i\frac{\phi_{0}-\phi_{1}}{2}}\ket{1}) \nonumber \\
 &\stackrel{\mbox{\tiny
BS2}}{\longrightarrow} & e^{i\frac{\phi_{1}+\phi_{2}}{2}}
(\cos\half(\phi_{0}-\phi_{1}) \ket{0} + i
\sin\half (\phi_{0}-\phi_{1}) \ket{1}),
\label{MZ}
\end{eqnarray}
where $\phi_0$ and $\phi_1$ are the settings of the two phase shifters and
the action of the beam-splitters is defined as
\begin{eqnarray}
  \ket{0}&{\longrightarrow} & \textstyle{\frac{1}{\sqrt 2}}
(\ket{0} +\ket{1}) \nonumber \\
  \ket{1}&{\longrightarrow} &\textstyle{\frac{1}{\sqrt 2}}
(\ket{0} -\ket{1})\;
\label{tran}
\end{eqnarray}
(and extends by linearity to states of the form $\alpha\ket{0} +
\beta\ket{1}$).
Here, we have ignored the $e^{i\frac{\phi_{0}+\phi_{0}}{2}}$ phase shift
in the reflected beam, which is irrelevant because the
interference pattern depends only on the {\em difference} between the phase
shifts in different arms of the interferometer.  The phase shifters in
the two paths can be tuned to effect any prescribed relative phase
shift $\phi = \phi_0 - \phi_1$ and to direct the particle with
probabilities $\cos^{2}\left(\frac{\phi}{2}\right)$ and
$\sin^{2}\left(\frac{\phi}{2}\right)$ respectively to detectors ``0"
and ``1".

The roles of the three key ingredients in this experiment are clear.
The first beam splitter prepares a superposition of possible paths,
the phase shifters modify quantum phases in different paths and the
second beam-splitter combines all the paths together.
As we shall see in the following sections, quantum algorithms follow this interferometry paradigm: a superposition
of computational paths is prepared by the Hadamard (or the Fourier)
transform, followed by a quantum function evaluation which effectively
introduces phase shifts into different computational paths, followed
by the Hadamard or the Fourier transform which acts somewhat in
reverse to the first Hadamard/Fourier transform and combines the
computational paths together.  To see this, let us start by
rephrasing Mach-Zehnder interferometry in terms of quantum
networks.

\section{Quantum gates \& networks}

\begin{figure}
\centering
\setlength{\unitlength}{1mm}
\begin{picture}(80,20)
\put(0,10){\line(1,0){7}}
\put(7,7){\framebox(6,6){\bf H}}
\put(52,7){\framebox(6,6){\bf H}}
\put(58,10){\line(1,0){10}}
\put(13,10){\line(1,0){9}}
\put(22,7){\framebox(21,6){$\phi=\phi_0-\phi_1$}}
\put(43,10){\line(1,0){9}}
\end{picture}
\label{fig-HPH}
  \caption{A quantum network composed of three single qubit gates.
  This network provides a hardware-independent description of any
  single-particle interference, including Mach-Zehnder
  interferometry.}
\end{figure}
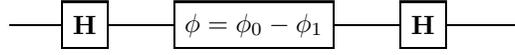

In order to avoid references to specific technological choices (hardware), let us
now describe our Mach-Zehnder interference experiment in more abstract
terms.  It is convenient to view this experiment as a {\em
quantum network} with three quantum logic gates (elementary
unitary transformations) operating on a qubit (a generic two-state
system with a prescribed computational basis $\{ \ket{0}, \ket{1}
\}$).  The beam-splitters will be now called the Hadamard gates and
the phase shifters the phase shift gates (see~Fig.~\ref{fig-HPH}).

The Hadamard gate is the single qubit gate $\bf H$ performing the
unitary transformation known as the Hadamard transform given by
(Eq.~\ref{tran})

\setlength{\unitlength}{0.030in}
\begin{equation}
  {\mbox{\bf H}}= \frac{1}{\sqrt 2}\left (
\begin{array}{cc}
  1 & 1 \\ 1& -1\\
\end{array}
\right ) \mbox{\hspace{3cm}} \mbox{
\begin{picture}(30,0)(15,15)
  \put(-4,14){$\ket{x}$} \put(5,15){\line(1,0){5}} \put(20,15){\line(1,0){5}}
  \put(10,10){\framebox(10,10){{\bf H}}}
\put(30,14){$\ket{0}+(-1)^x\ket{1}$}
\end{picture}
.}
\end{equation}

The matrix is written in the basis $\{\ket{0}, \ket{1}\}$ and the
diagram on the right provides a schematic representation of the gate
{\bf H} acting on a qubit in state $\ket{x}$, with $x=0,1$.  Using the
same notation we define the phase shift gate $\bf \phi$ as a single
qubit gate such that $\ket{0} \mapsto \ket{0}$ and $\ket{1}\mapsto
e^{i\phi}\ket{1}$,

\setlength{\unitlength}{0.030in}
\begin{equation}
  \phi = \left (
\begin{array}{cc}
  1 & 0 \\ 0& e^{i\phi}\\
\end{array}
\right ) \mbox{\hspace{3cm}} \mbox{
\begin{picture}(30,0)(15,15)
  \put(-4,14){$\ket{x}$} \put(5,15){\line(1,0){5}} \put(20,15){\line(1,0){5}}
  \put(10,10){\framebox(10,10){{$\phi$}}}
\put(30,14){$e^{ix\phi}\ket{x}$}
\end{picture}
.}
\end{equation}

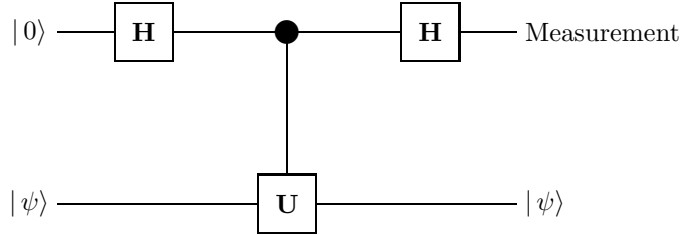
\begin{figure}
\centering
\begin{picture}(180,50)

\put(25,10){\makebox(0,0){$\ket{\psi}$}}
\put(25,40){\makebox(0,0){$\ket{0}$}}

\put(30,40){\line(1,0){10}}
\put(50,40){\line(1,0){40}}
\put(100,40){\line(1,0){10}}
\put(30,10){\line(1,0){35}}
\put(75,10){\line(1,0){35}}

\put(40,35){\framebox(10,10){\bf H}}
\put(90,35){\framebox(10,10){\bf H}}
\put(70,40){\circle*{4}}
\put(70,40){\line(0,-1){25}}
\put(65,5){\framebox(10,10){\bf U}}

\put(115,10){\makebox(0,0){$\ket{\psi}$}}
\put(125,40){\makebox(0,0){\mbox{Measurement}}}
\end{picture}
\caption{Phase factors can be introduced into different computational
paths via the controlled-$U$ operations.  The controlled-$U$ means
that the form of $U$ depends on the logical value of the control qubit
(the upper qubit).  Here, we apply the identity transformation to the
auxiliary (lower) qubits (i.e.  do nothing) when the control qubit is
in state $\ket{0}$ and apply a prescribed $U$ when the control qubit
is in state $\ket{1}$.  The auxiliary or the target qubit is initially
prepared in state $\ket{\psi}$ which is one of the eigenstates of $U$.}
\label{cphase}
\end{figure}

Let us explain now how the phase shift $\phi$ can be
``computed'' with the help of an auxiliary qubit (or a set of qubits)
in a prescribed state $\ket{\psi}$ and some controlled-$U$ transformation
where $U\ket{\psi}= e^{i\phi}\ket{\psi}$ (see Fig.~\ref{cphase}).

Here the controlled-$U$ is a transformation involving two qubits,
where the form of $U$ applied to the auxiliary or target qubit depends on the
logical value of the control qubit. For example, we can apply the
identity transformation to the auxiliary qubits (i.e.  do nothing)
when the control qubit is in state $\ket{0}$ and apply a prescribed
$U$ when the control qubit is in state $\ket{1}$.  In our example
shown in Fig.~\ref{cphase}, we obtain the following sequence of
transformations on the two qubits

\begin{eqnarray}
  \ket{0}\ket{\psi} \stackrel{H}{\longrightarrow}
  \textstyle{\frac{1}{\sqrt 2}}(\ket{0} + \ket{1})\ket{\psi} &
  \stackrel{c-U}{\longrightarrow} &
  \textstyle{\frac{1}{\sqrt 2}}(\ket{0} + e^{i\phi}\ket{1})
  \ket{\psi}\nonumber\\ &
  \stackrel{H}{\longrightarrow}& e^{(i\frac{\phi}{2})}
(\cos\textstyle{\phi \over 2}\ket{0}
 + i \sin\textstyle{\phi \over 2}\ket{1}) \ket{\psi}.
\label{sequ}
\end{eqnarray}

We note that the state of the auxiliary register $\ket{\psi}$, being an
eigenstate of $U$, is not altered along this network, but its
eigenvalue $e^{i\phi}$ is ``kicked back'' in front of the $\ket{1}$
component in the first qubit.  The sequence (\ref{sequ}) is equivalent to the steps of the Mach-Zehnder interferometer (\ref{MZ}) and, as was shown
in~\cite{CEMM}, the kernel of most known quantum algorithms.

\section{The first quantum algorithm}

Since quantum phases in interferometers can be introduced by some
controlled-$U$ operations, it is natural to ask whether effecting
these operations can be described as an interesting computational
problem.

Suppose an experimentalist, Alice, who runs the Mach-Zehnder
interferometer delegates the control of the phase shifters to her
colleague, Bob.  Bob is allowed to set up any value $\phi = \phi_0 - \phi_1$
and Alice's task is to estimate $\phi$.  Clearly for general $\phi$ this
involves running the device several times until Alice accumulates enough
data to estimate probabilities $P_{0}$ and $P_{1}$, however, if Bob
promises to set up $\phi$ either at $0$ or at $\pi$ then a
single-shot experiment can deliver the conclusive outcome (click in
detector ``0'' corresponds to $\phi=0$ and in detector ``1''
corresponds to $\phi=\pi$).  The first
quantum algorithm proposed by David Deutsch in 1985~\cite{Deutsch85}
is related to this effect.

We have seen in the previous section that a controlled-U
transformation can be used to produce a particular phase shift on the
control qubit corresponding to its eigenvalue on the auxiliary qubit.
If two eigenvalues of the controlled-U transformation lead to
different orthogonal states in the control qubit, a single measurement
on this qubit will suffice to distinguish the two cases.

For example consider the Boolean functions $f$ that map $\{0,1\}$ to
$\{0,1\}$.
There are exactly four such functions: two constant functions
($f(0)=f(1)=0$ and $f(0)=f(1)=1$) and two ``balanced'' functions
($f(0)=0, f(1)=1$ and $f(0)=1, f(1)=0$). It turns out that it is
possible to construct a controlled function evaluation such that two
possible eigenvalues are produced which may be used to determine
whether the function is constant or balanced. This is done in the
following way.

Let us formally define the operation of ``evaluating'' $f$ in terms of
the {\em $f$-controlled-NOT} operation on two bits: the first contains
the input value and the second contains the output value.  If the
second bit is initialised to $0$, the $f$-controlled-NOT maps $(x,0)$
to $(x,f(x))$.  This is clearly just a formalization of the operation
of computing $f$.  In order to make the operation reversible, the
mapping is defined for {\em all\/} initial settings of the two bits,
taking $(x,y)$ to $(x,y \oplus f(x))$, where $\oplus$ denotes addition
modulo two.

A single evaluation of the $f$-controlled-NOT on quantum
superpositions suffices to classify $f$ as constant or balanced. This
is the real advantage of the quantum method over the classical.
Classically if the $f$-controlled-NOT
operation may be performed only once then it is {\em impossible} to
distinguish between
balanced and constant functions.  Whatever the
outcome, both possibilities (balanced and constant) remain for $f$.
This corresponds to our classical intuition about the problem since it
involves determining not particular values of $f(0)$ and $f(1)$, but a
global property of $f$. Classically  to
determine this global property of $f$, we have to evaluate both $f(0)$
and $f(1)$, which involves evaluating $f$ twice.

\begin{figure}
\centering
\begin{picture}(180,50)

\put(18,10){\makebox(0,0){$\ket{0}-\ket{1}$}}
\put(25,40){\makebox(0,0){$\ket{0}$}}

\put(30,40){\line(1,0){10}}
\put(50,40){\line(1,0){40}}
\put(100,40){\line(1,0){10}}
\put(30,10){\line(1,0){35}}
\put(75,10){\line(1,0){35}}

\put(40,35){\framebox(10,10){\bf $H$}}
\put(90,35){\framebox(10,10){\bf $H$}}
\put(70,40){\circle*{4}}
\put(70,40){\line(0,-1){25}}
\put(65,5){\framebox(10,10){\bf $U_f$}}

\put(122,10){\makebox(0,0){$\ket{0}-\ket{1}$}}
\put(125,40){\makebox(0,0){\mbox{Measurement}}}
\end{picture}
\caption{Quantum network which implements Deutsch's algorithm. The
middle gate is the $f$-controlled-NOT which evaluates one of the four
functions $f:\; \{0,1\}\mapsto \{0,1\}$. If the first qubit is
measured to be $\ket{0}$, then the function is constant, and if
$\ket{1}$, the function is balanced.}
\label{deutsch}
\end{figure}
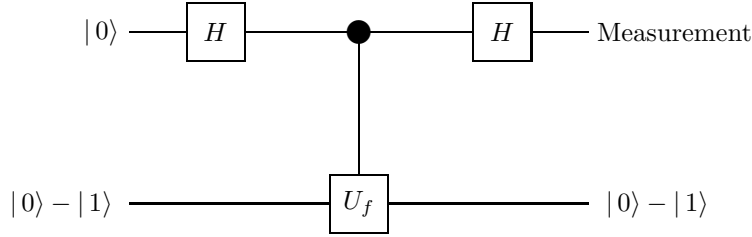

Deutsch's quantum algorithm has the same mathematical structure as the
Mach-Zehnder interferometer, with the two phase settings $\phi = 0, \pi$.
It is best represented as
the quantum network shown in Fig.~\ref{deutsch}, where the middle
operation is the $f$-controlled-NOT, which can be defined as:

\begin{equation}
  \ket{x}\ket{y} \stackrel{f-c-N}{\longrightarrow} \ket{x}\ket{y
    \oplus f(x)}\;.
\label{eqf}
\end{equation}

The initial state of the qubits in the quantum network is
$\ket{0}(\ket{0}-\ket{1})$ (apart from a normalization factor, which
will be omitted in the following).  After the first Hadamard
transform, the state of the two qubits has the form
$(\ket{0}+\ket{1})(\ket{0}-\ket{1})$.  To determine the effect of the
$f$-controlled-NOT on this state, first note that, for each $x \in
\{0,1\}$,
\begin{equation}
  \ket{x}(\ket{0}-\ket{1}) \stackrel{f-c-N}{\longrightarrow}
  \ket{x}(\ket{0\oplus f(x)}-\ket{1\oplus f(x)})=(-1)^{f(x)}
  \ket{x}(\ket{0}-\ket{1}) \;.
\end{equation}
Therefore, the state after the $f$-controlled-NOT is
\begin{equation}
  ((-1)^{f(0)}\ket{0}+(-1)^{f(1)}\ket{1}) (\ket{0}-\ket{1})\;.
\label{st-d}
\end{equation}
That is, for each $x$, the $\ket{x}$ term acquires a phase factor of
$(-1)^{f(x)}$,
which corresponds to the eigenvalue of the state of the auxiliary qubit under
the action of the operator that sends $\ket{y}$ to $\ket{y \oplus f(x)}$.

This state can also be written as
\begin{equation}
(-1)^{f(0)}(\ket{0}+(-1)^{f(0)\oplus f(1)}\ket{1})(\ket{0}-\ket{1})\;,
\end{equation}
which, after applying the second Hadamard transform to the first
qubit, becomes
\begin{equation}
(-1)^{f(0)}\ket{f(0) \oplus f(1)} (\ket{0}-\ket{1})\;.
\end{equation}
Therefore, the first qubit is finally in state $\ket{0}$ if the
function $f$ is constant and in state $\ket{1}$ if the function is
balanced, and a measurement of this qubit distinguishes these cases
with certainty.

The Mach-Zehnder interferometer with phases $\phi_0$ and $\phi_1$ each
set to either 0 or $\pi$ can be regarded as an implementation of the above
algorithm.
In this case, $\phi_0$ and $\phi_1$ respectively encode $f(0)$ and $f(1)$
(with $\pi$ representing 1), and a single photon can query both phase
shifters (i.e.\ $f(0)$ and $f(1)$) in superposition.
More recently, this algorithm (Fig. 4) has been implemented using a very
different quantum physical technology, nuclear magnetic resonance
\cite{Jones, Chuang}.

More general algorithms may operate not just on single qubits, as in
Deutsch's case, but on sets of qubits or `registers'. The second qubit
becomes an auxiliary register $\ket{\psi}$ prepared in a superposition
of basis states, each weighted by a different
phase factor,
\begin{equation}
\ket{\psi}=\sum_{y=0}^{2^m - 1} e^{-2\pi i y/2^m}\ket{y}.
\label{reg}
\end{equation}
In general, the middle gate which produces the phase shift
is some controlled function evaluation. A controlled function evaluation
operates on its second input, the `target', according to the state of the
first input, the `control'. A controlled function $f$ applied to a
control state $\ket{x}$, and a target state $\ket{\psi}$ gives
\begin{equation}
\ket{x}\ket{\psi} \longrightarrow \ket{x}\ket{\psi + f(x)}.
\end{equation}
where the addition is mod $2^m$.
Hence for the register in state~(\ref{reg})
\begin{equation}
\ket{x} \sum_{y=0}^{2^m - 1} e^{-2\pi i y/ 2^m}\ket{y}
\longrightarrow e^{2\pi i
  f(x)/2^m}\ket{x} \sum_{y=0}^{2^m - 1}
e^{-2\pi i (y+f(x))/2^m}\ket{y+f(x)}
=e^{2\pi i f(x)/2^m}\ket{x}\ket{\psi}.
\end{equation}
Effectively a phase shift proportional to the value of $f(x)$ is
produced on the first input.

We will now see how phase estimation on
registers may be carried out by networks consisting of only two types
of quantum gates: the Hadamard gate  $\bf H$ and the
conditional phase shift $\bf R(\phi)$. The conditional phase shift is the
two-qubit gate $\bf R(\phi)$ defined as
\begin{equation}
  {\mbox{\bf R}}(\phi)=\left.\left (
  \begin{array}{cccc}
    1 & 0 & 0 & 0\\ 0 & 1 & 0 & 0\\ 0 & 0 & 1 & 0\\ 0 & 0 & 0 &
    e^{i\phi}\\
  \end{array}
\right )
\mbox{\hspace{1.5cm}}
\mbox{
\begin{picture}(25,0)(0,20)
  \put(-4,14){$\ket{y}$} \put(-4,29){$\ket{x}$} \put(5,15){\line(1,0){20}}
\put(5,30){\line(1,0){20}}
  \put(15,30){\circle*{3}} \put(15,15){\line(0,1){15}}
%  \put(10,10){\framebox(10,10){$\phi$}}
\put(15,15){\circle*{3}}
\end{picture}}
\quad \right \} e^{i xy \phi}\ket{x}\ket{y}.
\end{equation}
The matrix is written in the basis $\{\ket{0}\ket{0}, \ket{0}\ket{1},
\ket{1}\ket{0}, \ket{1}\ket{1}\}$, (the diagram on the right shows the
structure
of the gate). For some of the known quantum algorithms, when working with
registers, the Hadamard transformation,
corresponding to the beamsplitters in the interferometer, is
generalised to a quantum Fourier transform.

\section{Quantum Fourier transform and computing phase shifts}

The discrete Fourier transform is a unitary transformation of a
$s$--dimensional vector
\begin{equation}
(f(0), f(1), f(2),\ldots,f(s-1))
\rightarrow (\tilde f(0),\tilde f(1),\tilde f(2),\ldots,
\tilde f(s-1))
\end{equation}
defined by:
\begin{equation}
   \tilde f(y)=\frac{1}{\sqrt s}\sum_{x=0}^{s-1}e^{2\pi i xy/s} f(x),
   \label{def}
\end{equation}
where $f(x)$ and $\tilde f(y)$ are in general complex numbers. In the
following, we assume that $s$ is a power of $2$, i.e., $s=2^n$
for some $n$; this is a natural choice when binary coding is used.

The quantum version of the discrete Fourier transform (QFT) is a unitary
transformation which can be written in a chosen computational basis
$\{ | 0 \rangle,| 1 \rangle,\ldots,| 2^{n}-1 \rangle\}$
as,
\begin{equation}
   | x \rangle \longmapsto \frac{1}{\sqrt{s}}
   \sum_{y=0}^{s-1}
   \exp(2\pi ixy/s)\: | y \rangle .
\label{qftdef}
\end{equation}
More generally, the $\mbox{QFT}$ effects the discrete Fourier transform
of the input amplitudes. If
\begin{equation}
   \mbox{\rm QFT}: \sum_x f(x)| x \rangle \longmapsto
   \sum_y \tilde{f}(y)| y \rangle,
\end{equation}
then the coefficients $\tilde{f}(y)$ are the discrete Fourier
transforms of the $f(x)$'s.

A given phase $\phi_x = 2\pi x / 2^n$ can be encoded by a QFT. In this
process the information about $\phi_x$ is distributed between states
of a register. Let $x$ be represented in binary as
$x_0 \ldots x_{n-1} \in \{0,1\}^n$,
where $x=\sum_{i=0}^{n-1}x_{i}2^{i}$ (and similarly for $y$). An important
 observation is that the QFT of $x$, $\sum_{y=0}^{s-1}
\exp(2\pi ixy/s)\: | y \rangle$, is
unentangled, and can in fact be factorised as
\begin{equation}
  (\ket{0} + e^{i\phi_{x}}\ket{1}) (\ket{0} + e^{i2\phi_{x}}\ket{1})
  \cdots (\ket{0} + e^{i2^{n-1}\phi_{x}}\ket{1})\; .
\label{ft2}
\end{equation}
The network for performing the QFT is shown in Fig.~\ref{fig-QFT}. The
input qubits are initially in some state $\ket{x} =
\ket{x_0}\ket{x_1}\ket{x_2}\ket{x_3}$ where $x_0x_1x_2x_3$ is the
binary representation of $x$, that is, $x=\sum_{i=0}^{3}x_i2^i$. As the number of qubits becomes large, the rotations $R(\pi /2^n)$
will require exponential precision, which is impractical. Fortunately,
the algorithm will work even if we omit the small rotations,
\cite{Coppersmith, BEST}.

The general case of $n$ qubits requires a simple extension of the network
following the same pattern of ${\bf H}$ and
${\bf R}$ gates.

\setlength{\unitlength}{0.02in}
\begin{figure}
\centering
\begin{picture}(210,120)(40,0)

\put(0,34){\makebox(20,12){$\ket{x_3}$}}
\put(0,54){\makebox(20,12){$\ket{x_2}$}}
\put(0,74){\makebox(20,12){$\ket{x_1}$}}
\put(0,94){\makebox(20,12){$\ket{x_0}$}}

\put(240,34){\makebox(20,12){$\ket{0}+e^{2\pi ix/2^4}\ket{1}$}}
\put(240,54){\makebox(20,12){$\ket{0}+e^{2\pi ix/2^3}\ket{1}$}}
\put(240,74){\makebox(20,12){$\ket{0}+e^{2\pi ix/2^2}\ket{1}$}}
\put(240,94){\makebox(20,12){$\ket{0}+e^{2\pi ix/2}\ket{1}$}}

\put(43,16){$\bf H\;R(\pi)\;H\;R(\pi/2)R(\pi)\;H\;R(\pi/4)R(\pi/2)R(\pi)\;H$}

\put(20,40){\line(1,0){170}}
\put(190,34){\framebox(12,12){$H$}}
\put(202,40){\line(1,0){18}}
\put(20,60){\line(1,0){104}}
\put(124,54){\framebox(12,12){$H$}}
\put(136,60){\line(1,0){84}}
\put(20,80){\line(1,0){54}}
\put(74,74){\framebox(12,12){$H$}}
\put(86,80){\line(1,0){134}}
\put(20,100){\line(1,0){14}}
\put(34,94){\framebox(12,12){$H$}}
\put(46,100){\line(1,0){174}}

\put(60,100){\circle*{4}}
\put(60,100){\line(0,-1){20}}
\put(60,80){\circle*{4}}

\put(100,100){\circle*{4}}
\put(100,60){\circle*{4}}
\put(100,60){\line(0,1){40}}

\put(110,80){\circle*{4}}
\put(110,60){\circle*{4}}
\put(110,60){\line(0,1){20}}

\put(150,40){\circle*{4}}
\put(160,40){\circle*{4}}
\put(170,40){\circle*{4}}
\put(150,40){\line(0,1){60}}
\put(160,40){\line(0,1){40}}
\put(170,40){\line(0,1){20}}
\put(150,100){\circle*{4}}
\put(160,80){\circle*{4}}
\put(170,60){\circle*{4}}

\end{picture}
\caption{The quantum Fourier transform (QFT) network operating on
four qubits.  If the input state represents number $x=\sum_k 2^kx_k$
the output state of each qubit is of the form
$\ket{0}+e^{i2^{{n-1-k}}\phi_x}\ket{1}$, where $\phi_x=2\pi x/2^n$ and
$k=0,1,2\ldots n-1$.  N.B. there are three different types of the
$R(\phi)$ gate in the network above: $R(\pi)$, $R(\pi/2)$ and
$R(\pi/4)$. The size of the rotation is indicated by the distance
between the `wires'.}
\label{fig-QFT}
\end{figure}
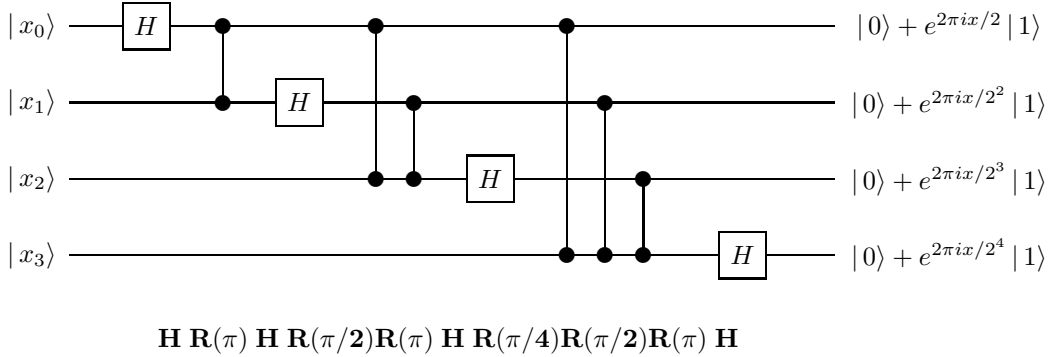

States of the form~(\ref{ft2}) are produced by function evaluation in a
quantum computer. Suppose that $U$ is any unitary transformation on $m$
qubits and
$\ket{\psi}$ is an eigenvector of $U$ with eigenvalue $e^{i\phi}$. The
scenario is that we do not explicitly know $U$
or $\ket{\psi}$ or $e^{i \phi}$, but instead are given devices that
perform controlled-$U$, controlled-$U^{2^1}$, controlled-$U^{2^2}$
and so on until we reach controlled-$U^{2^{n-1}}$.  Also, assume
that we are given a single preparation of the state $\ket{\psi}$. From this, our goal is to obtain an $n$-bit estimator of $\phi$.

In a quantum algorithm a quantum
state of the form
\begin{equation}
(\ket{0} + e^{i 2^{n-1} \phi}\ket{1})
(\ket{0} + e^{i 2^{n-2} \phi}\ket{1})
\cdots
(\ket{0} + e^{i \phi}\ket{1})
\label{qftphi}
\end{equation}
is created by applying the network of Fig.~\ref{fs}.

Then, in the special case where $\phi = 2\pi
x/2^{n}$, the state $\ket{x_0\cdots x_{n-1}}$ (and hence $\phi$) can be
obtained by just applying the inverse of the QFT (which is the network
of Fig.~\ref{fig-QFT} in the backwards direction and with the qubits in
reverse order). If $x$ is an $n$-bit
number this will produce the state $\ket{x_{0}\cdots x_{n-1}}$ exactly
(and hence the exact value $\phi$).

However, $\phi$ is not in general a fraction of a power of two (and
may not even be a rational number).  For such a $\phi = 2\pi \omega$,
it turns out
that applying the inverse of the QFT produces the best $n$-bit
approximation of $\omega$ with probability at least $4 / \pi^2 \approx
0.41 $~\cite{CEMM}.  The probability of obtaining the
best\footnote{Though this process produces the best estimate
of $\omega$ with significant probability, it is not necessarily the best
estimator of $\omega$, since, for example, we might be able to
to obtain as close an estimate with higher probability.
See \cite{DDEMM} for details.}
estimate can be made $1-\delta$ for any $\delta$, $ 0< \delta < 1$,
by creating the state in equation
(\ref{qftphi}) but with $n + O(\log(1/\delta))$ qubits and
rounding the answer off to the nearest $n$ bits \cite{CEMM}.

\setlength{\unitlength}{0.03cm}
\begin{figure}
\centering
\begin{picture}(290,180)(105,380)
\thicklines
\put(160,380){\framebox(20,60){$U^{2^{0}}$}}
\put(220,380){\framebox(20,60){$U^{2^{1}}$}}
\put(285,380){\framebox(20,60){$U^{2^{2}}$}}
\put(110,425){\line( 1, 0){ 50}}
\put(180,425){\line( 1, 0){ 40}}
\put(240,425){\line( 1, 0){ 45}}
\put(305,425){\line( 1, 0){ 80}}
\put(110,415){\line( 1, 0){ 50}}
\put(180,415){\line( 1, 0){ 40}}
\put(240,415){\line( 1, 0){ 45}}
\put(305,415){\line( 1, 0){ 80}}
\put(110,405){\line( 1, 0){ 50}}
\put(180,405){\line( 1, 0){ 40}}
\put(240,405){\line( 1, 0){ 45}}
\put(305,405){\line( 1, 0){ 80}}
\put(110,395){\line( 1, 0){ 50}}
\put(180,395){\line( 1, 0){ 40}}
\put(240,395){\line( 1, 0){ 45}}
\put(305,395){\line( 1, 0){ 80}}
\put(105,560){\line( 1, 0){280}}
\put(105,525){\line( 1, 0){280}}
\put(105,485){\line( 1, 0){280}}
\put(170,440){\line( 0, 1){ 45}}
\put(230,440){\line( 0, 1){ 85}}
\put(295,440){\line( 0, 1){120}}
\put(80,407){$\ket{\Phi}$}
\put(397,407){$\ket{\Phi}$}
\put(50,485){$\ket{0}+\ket{1}$}
\put(50,525){$\ket{0}+\ket{1}$}
\put(50,560){$\ket{0}+\ket{1}$}
\put(393,485){$\ket{0}+e^{i2^{0}\phi}\ket{1}$}
\put(393,525){$\ket{0}+e^{i2^{1}\phi}\ket{1}$}
\put(393,560){$\ket{0}+e^{i2^{2}\phi}\ket{1}$}
\put(170,485){\circle*{6}}
\put(230,525){\circle*{6}}
\put(295,560){\circle*{6}}
\end{picture}
\caption{The network which computes phase shifts in Shor's algorithms;
it also implements the modular exponentiation function via repeated
squarings.}
\label{fs}
\end{figure}
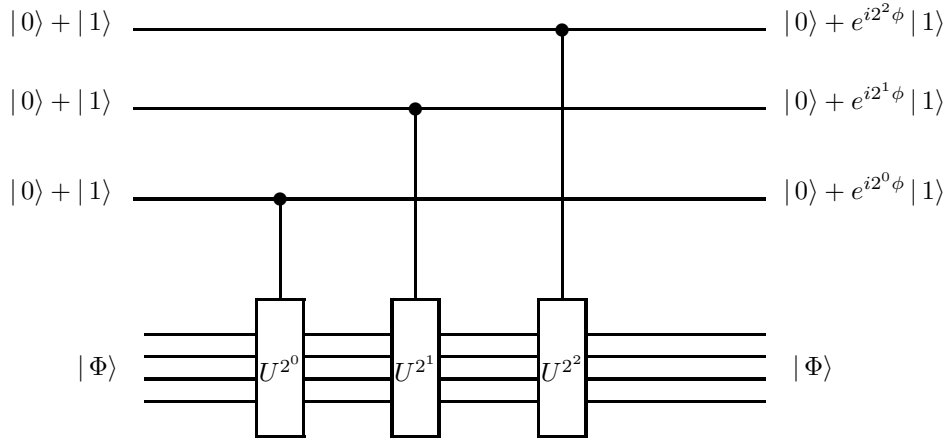

\section{Examples}

We will now illustrate the general framework described in the
preceding section by showing how some of the most important quantum
algorithms can be viewed in this light. We start with Shor's quantum
algorithm for efficient factorisation (for a comprehensive discussion of
quantum factoring
see~\cite{Shor,EJ, CEMM}).

\subsection{Quantum Factoring}
Shor's quantum factoring of an integer $N$ is based on calculating
the period of the function $f(x) = a^x\bmod N$ for a randomly
selected integer $a$ between $1$ and $N$. For any positive integer
$y$, we define $y \bmod N$ to be the remainder (between $0$ and
$N-1$) when we divide $y$ by $N$. More generally, $y \bmod N$ is
the unique positive integer $\overline{y}$ between $0$ and $N-1$
such that $N$ evenly divides $y - \overline{y}$.  For example, $2
\bmod 35 = 2$, $107 \bmod 35 = 2$, and $-3 \bmod 35 = 32$. We can
test if $a$ is relatively prime to $N$ using the Euclidean
algorithm. If it is not, we can compute the greatest common divisor
of $a$ and $N$ using the extended Euclidean algorithm. This will
factor $N$ into two factors $N_1$ and $N_2$ (this is called
\emph{splitting} $N$). We can then test if $N_1$ and $N_2$ are
powers of primes, and otherwise proceed to split them if they are
composite.  We will require at most $\log_2(N)$ splittings before
we factor $N$ into its prime factors. These techniques are
summarised in \cite{MOV}.

It turns out that for increasing
powers of $a$, the remainders form a repeating sequence with a
period $r$.  We can also call $r$ the
\emph{order} of $a$ since $a^r = 1 \bmod N$.
Once $r$ is known,
factors of $N$
are obtained
by calculating the greatest common divisor
of $N$ and
$a^{r/2}\pm 1$.

Suppose we want to factor $35$ using this method.  Let $a=4$.  For
increasing $x$ the function $4^x\bmod 35$ forms a repeating sequence
$4,16,32,29,9,1,4,16,29,32,9,1,\ldots$.
The period is $r=6$, and $a^{r/2}\bmod 35
=29$.  Then we take the greatest common divisor of $28$ and $35$, and
of $30$ and $35$, which gives us $7$ and $5$, respectively, the two
factors of $35$.  Classically, calculating $r$ is at least as difficult
as trying to factor $N$; the execution time of the
%RC
best currently-known algorithms
grows exponentially with the number of digits in $N$. Quantum computers can
find $r$ very efficiently.

Consider the unitary transformation
$U_a$ that maps $\ket{x}$ to $\ket{a x \bmod N}$.
Such a transformation is realised by simply implementing
the reversible classical network for multiplication by
$a$ modulo $N$ using quantum gates.
The transformation $U_a$, like the element $a$, has
order $r$, that is, $U_a^r = I$, the identity operator.
Such an operator has eigenvalues of the form $e^{2 \pi i k \over r}$
for $k = 0,1,2, \ldots, r-1$.
In order to formulate Shor's algorithm in terms of phase
estimation let us apply the construction from the last section taking
\begin{equation}
\ket{\psi} = \sum_{j=0}^{r-1} e^{-2 \pi i j \over r}
\ket{a^j \mbox{ mod } N }\;.
\end{equation}

Note that
$\ket{\psi}$ is an eigenvector of $U_a$ with eigenvalue $e^{2\pi i ({1
\over r})}$.  Also, for any $j$, it is possible to implement
efficiently a controlled-$U_a^{2^j}$ gate by a sequence of squaring
(since $U_a^{2^j} = U_{a^{2^j}}$).
Thus, using the state
$\ket{\psi}$ and the implementation of controlled-$U_a^{2^j}$ gates, we
can directly apply the method of the last section to efficiently
obtain an estimator of ${1 \over r}$.

The problem with the above method is that we are aware of
no straightforward efficient method to prepare state $\ket{\psi}$,
however, let us notice that almost any state $\ket{\psi_{k}}$ of the form
\begin{equation}
\ket{\psi_{k}} =
\sum_{j=0}^{r-1} e^{-\frac{2\pi i k j}{r}}\ket{a^j \bmod N}\;,
\end{equation}
where $k$ is from $\{0,\ldots,r-1\}$ would also do the job. For
each $k \in \{0,1,\ldots,r-1\}$, the eigenvalue of state
$\ket{\psi_{k}}$ is $e^{2 \pi i ({k \over r})}$.  We can again use
the technique from the last section to efficiently determine ${k
\over r}$ and if $k$ and $r$ are coprime then this yields
\footnote{If the estimate $y/2^m$ of $k/r$ satisfies
\[ \left| {y \over 2^m} - {k \over r} \right|  < {1 \over 2N^2}, \]
then there is a unique rational of the form ${a \over b}$
with $0 < b \leq N$ satisfying
\[ \left| {y \over 2^m} - {a \over b} \right| < {1 \over 2N^2}.\]
Consequently, $a/b = k/r$, and the continued fractions algorithm
will find the fraction for us.  We might be unlucky and get a $k$
like $0$, but with even $2$ repetitions with random $k$ we can find
$r$ with probability at least $0.54$ \cite{CEMM}.} $r$.
Now the key
observation is that
\begin{equation}  \label{sum.of.eigen}
\ket{1} = \sum_{k=1}^{r} \ket{\psi_k}\;,
\end{equation}
and $\ket{1}$ {\em is} an easy state to prepare.

If we substituted $\ket{1}$ in place of $\ket{\psi}$ in the last
section then effectively we would be estimating one of the $r$,
randomly chosen, eigenvalues $e^{2 \pi i ({k \over r})}$.  This
demonstrates that Shor's algorithm, in effect, estimates the
eigenvalue corresponding to an eigenstate of the operation $U_a$ that
maps $\ket{x}$ to $\ket{ax \bmod N}$.
A classical
procedure - the continued fractions algorithm - can be employed
to estimate $r$ from these results. The value of $r$ is then used to
factorise the integer.

\subsection{Finding hidden subgroups}

\begin{figure}
\centering
\input{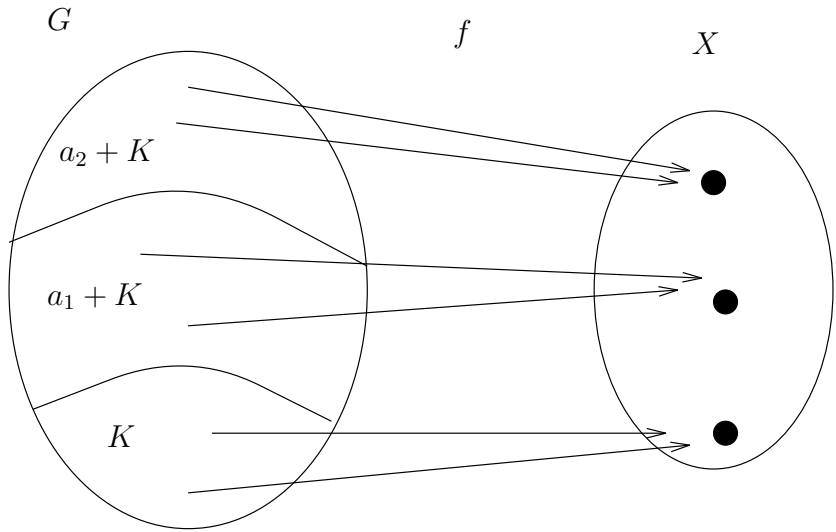}
\caption{A function $f$ mapping elements of a group $G$ to
a set $X$ with a hidden subgroup $K$.  This means that $f(g_1) = f(g_2)$
 if and only if $g_1$ and $g_2$ are in the same coset of $K$.}
\label{hidden.sub}
\end{figure}

A number of algorithms can be generalised in terms of group theory as
examples of finding hidden subgroups. For any $g \in G$, the coset $gK$, of the subgroup $K$ is defined as $\{gK | g\in G\}$.
Say we have a function $f$ which maps a group $G$ to a set $X$,
and $f$ is constant on each coset of the subgroup $K$,
and distinct on each coset, as illustrated in
Figure \ref{hidden.sub}.
\setlength{\unitlength}{0.03cm}
In other words, $f(x) = f(y)$ if and only if $x-y$ is
an element of $K$.

In Deutsch's case, $G = \{0,1\}$ with addition mod $2$ as the group operation,
and $X$ is also $\{0,1\}$. There are two possible subgroups $K$:
$\ket{0}$, and $G$ itself. We are given a black-box $U_f$ for computing $f$
\[ \ket{x}\ket{y} \rightarrow \ket{x}\ket{y \oplus f(x) }. \]
There are two cosets of the subgroup $\{0\}$: $\{0\}$ and
$\{1\}$. If the function is defined to be constant and distinct on
each coset, it must be balanced. On the other hand, there is only one coset
of the other subgroup $G$, the group itself. In this case the function is
constant. With our specially chosen eigenvector $\ket{0} - \ket{1}$
the algorithm always outputs $\ket{0}$ if $K = \{0,1\}$
($f$ is constant),
and $\ket{1}$ if $K = \{0\}$, (f is balanced). Therefore we can view
Deutsch's algorithm as distinguishing between the `hidden subgroups'.

The hidden subgroup problem also encompasses the
problem of finding orders of elements in a group,
of which the factoring algorithm is a special case.
In quantum factoring, we wish to
find the order $r$ of the element $a$ in some group represented by $X$.
Here $G$ is the group of integers $\mathbf{Z}$ and
$K$ is the additive subgroup $r \mathbf{Z}$
of integer multiples of $r$, where $r$ is the order of $a$, and $a$ is
from the multiplicative group of integers modulo $N$.
The function $f$ maps $x$ to $a^x$ mod $N$.

\begin{figure}
\centering
\input{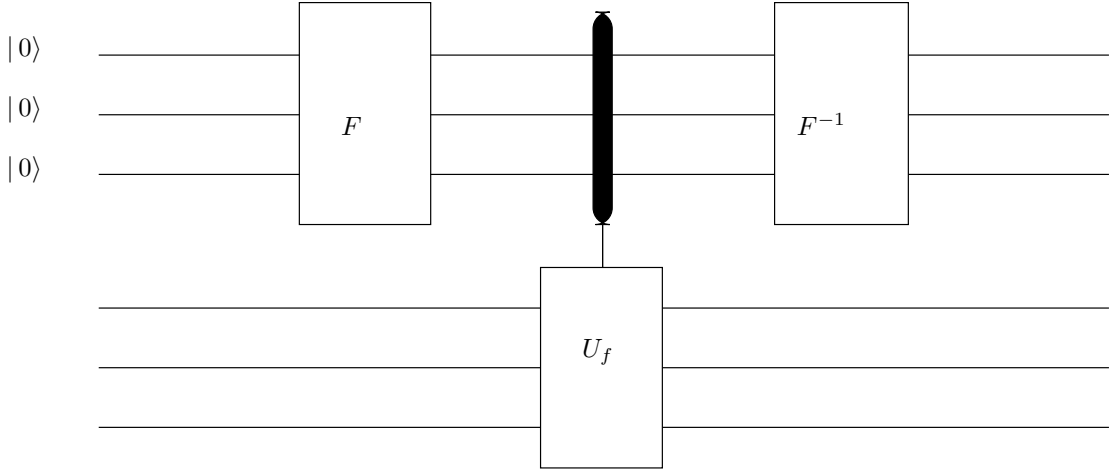}
\caption{The generic structure of a quantum network
solving any instance of the hidden subgroup problem.
The first register contains tuples of integers
corresponding to the Abelian group $G$.
The role of the first Fourier transform is to create
a superposition of many computational paths corresponding
to different elements of $G$.
The evaluation of the function simply kicks phases
back into the control register states, and the
final inverse Fourier transform produces the estimates
of the eigenvalues of operators related to the function $f$.
The set of eigenvalues corresponding to a particular eigenvector
produces an element orthogonal to $K$.  By collecting enough
such orthogonal elements we can efficiently find a generating set for $K$.}
\label{hidden.network}
\end{figure}

The output $\ket{y}$ in this case estimates an element
which is orthogonal
\footnote{By \emph{orthogonal} here,
we are not referring to the orthogonality of states in our
computational Hilbert space.  When we say $k/r$ is orthogonal to $K
= r\mathbf{Z}$, we mean that $\mbox{exp}(2\pi i z \frac{k}{r}) = 1$
for every $z \in K$.  This notion of orthogonality generalises to
groups with several generators as well.  } to the subgroup $K$. The
output $\ket{z}$ corresponds to the estimate $z/2^n$  of the
eigenvalue $k/r$ of the operator $U_a$ which maps $\ket{x}$ to
$\ket{ax}$ (that is, the operator which maps $\ket{f(g)}$ to
$\ket{f(g+1)}$) on the eigenvector $\ket{\psi_k}$. In general, for
any function $f$ mapping a finitely generated Abelian group $G$ to
a finite set $X$, the quantum network shown in figure
\ref{hidden.network} will output an estimate of a random element
orthogonal to the hidden subgroup $K$. With enough such elements,
we can easily determine $K$ using linear algebra.

By framing algorithms in terms of
hidden subgroups, it may be possible to think of other problems
associated with this structure in groups which we can treat with
quantum algorithms. A number of algorithms have already been cast in
this language, including Deutsch's problem \cite{Deutsch85, CEMM},
Simon's problem \cite{Simon}, factoring integers \cite{Shor}, finding
discrete logarithms \cite{Shor}, Abelian stabilisers \cite{Kitaev},
self-shift-equivalences \cite{Grigoriev},
and others \cite{BL} (see \cite{Mosca.Ekert}
and \cite{Hoyer} for details).

\subsection{Quantum Counting and Searching}

The first quantum algorithm for searching was constructed by
Grover~\cite{Grover}. This has led to
a large class of searching and counting algorithms.

We again consider a function $f$, this time mapping
us from a set $X$ to the set $\{0,1\}$.

We might wish to decide if there is a solution to
$f(x)=1$ ( the \emph{decision} problem) , or to
actually find a solution to $f(x)=1$ (the \emph{searching} problem).
We might be more demanding and want to know
how many solutions $x$ there are to
$f(x)=1$ (\emph{counting problem}).
Small cases of the searching \cite{GCK, JHM}
and counting \cite{JM}
algorithms have been
implemented using NMR technology.

In this section we will show how approximate
quantum counting can easily be phrased
as an instance of phase estimation, and quantum searching
as an instance of inducing a desired relative phase
between two eigenvectors.

In the following sections analysing quantum counting
and searching, we will be considering the \emph{Grover iterate}

\begin{equation}
\it{G} = -A U_0 A^{-1} U_f
\end{equation}
which was defined in
\cite{Grover} with $A$ as the Hadamard transform.
It was later generalised
in \cite{BH}, \cite{Grover97}, \cite{BHT} and \cite{BBHT}
with $A$ being any transformation
%\[ \ket{0} \rightarrow \ket{u} \] where $\ket{u} =
such that $A\ket{0}$ contains
a solution to $f(x)=1$ with non-zero amplitude, i.e.\
$|\bra{x}A\ket{0}|^2 > 0$ for some $x$ with $f(x)=1$.
The operator $U_f$ maps
\[ \ket{x} \rightarrow -\ket{x} \] for all $x$ satisfying $f(x)=1$,
and the operator $U_0$ maps
\[ \ket{0} \rightarrow -\ket{0} \]
leaving the remaining basis states alone.
Note that this $U_f$ is slightly different than the
standard $U_f$ which maps $\ket{x}\ket{b}$ to $\ket{x}\ket{b \oplus f(x)}$,
but can be easily obtained from it
by setting $\ket{b}$ to $\ket{0} - \ket{1}$.
%we can realise the $

\subsubsection{Quantum Counting}

Quantum counting was first discussed in \cite{BBHT}, where
it was observed that the Grover iterate is almost periodic with
a period dependent on the number of solutions. Therefore the
techniques of period-finding, as in Shor's algorithm, were applied \cite{BHT}. It
is also possible to think of the problem as a phase estimation (see
\cite{Mosca.Brno}).

We simply observe that the eigenvalues\footnote{The eigenvalue $-1$ has
multiplicity $j-1$, $1$ has multiplicity
$N-j-1$, and
$e^{2 \pi i \omega}$ and
$e^{-2 \pi i \omega}$ each have multiplicity $1$.
If $j = 0$, then $1$ has multiplicity $N$, (note that
$e^{2\pi i \omega_0}
= e^{-2\pi i \omega_0} = 1$),
if $j = N$, then $-1$ has multiplicity $N$, ($e^{2\pi i \omega_N}
= e^{-2\pi i \omega_N} = -1$).}
of $G$ are $1$, $-1$,
$e^{2 \pi i \omega_j}$,
and $e^{-2 \pi i \omega_j}$  where $f(x) = 1$ has $j$ solutions
and
\[ e^{2 \pi i \omega_j} = 1 - 2j/N + 2i\sqrt{j/N - (j/N)^2} .\]
Let $X_1$ denote the set of solutions to $f(x)=1$, and $X_0$ denote
the set of solutions to $f(x)=0$.
Estimating $\omega_j$ (or $-\omega_j$) will give us information about
the number of solutions to $f(x) = 1$.  For example,
for small $\omega_j$, the number of solutions, $j$, is
roughly $ N \pi^2 \omega_j^2$ since $\cos(2\pi \omega_j) = 1 -2j/N
\approx 1 - 2 \pi^2 \omega_j^2$
for small $\omega_j$.

We can use the techniques of the previous sections to
estimate this phase $\omega_j$ provided we know how to create
a starting state containing the eigenvectors with eigenvalues
$e^{2 \pi i \omega_j}$ and
$e^{-2 \pi i \omega_j}$.  For non-trivial $j$,
these eigenvectors are given by
\begin{eqnarray}
&&
\ket{\psi_+} = \frac{1}{\sqrt{2}}(\ket{X_1} + i \ket{X_0}) \\
&&
\ket{\psi_-} = \frac{1}{\sqrt{2}}(\ket{X_1} - i \ket{X_0})
\end{eqnarray}
where
\begin{eqnarray}
&&
\ket{X_1} = \frac{1}{\sqrt{j}}\sum_{f(x)=1} \ket{x} \\
&&
\ket{X_0} = \frac{1}{\sqrt{N-j}}\sum_{f(x)=0} \ket{x}.
\end{eqnarray}

Fortunately, the starting state
\[ A \ket{0} = \frac{1}{\sqrt{N}}\sum_{x=0}^{N-1} \ket{x}  \]
is equal to
\begin{equation}
\frac{1}{\sqrt{2}}(e^{-2 \pi i \theta_j} \ket{\psi_+} +
e^{2 \pi i \theta_j} \ket{\psi_-})
\end{equation}
for some real number $\theta_j$, which is not important
as far as counting is concerned, since all that is required for the
phase estimation procedure is any superposition of these two eigenvectors of
$G$.

Thus using a controlled-$G$,
controlled-$G^2$,
\ldots, and a controlled-$G^{2^n}$, (as done
with controlled-$U$s in Figure~\ref{fs})  and applying a quantum
Fourier transform, we can get an $n$-bit estimate of
either $\omega_j$ or $-\omega_j$.  This gives us an estimate of $j$,
the number of solutions.
Note that, unlike in the case of finding orders, there are
in general no short-cuts for computing higher powers of $G$.
That is, computing $G^{2^n}$ requires $2^n$ repetitions of $G$.
%The one qubit functions $U_{f(0)}$ and $U_{f(1)}$ which
%map $\ket{y}$ to
%$\ket{y\oplus f(0)}$ and
%$\ket{y\oplus f(1)}$ respectively, have eigenvalues
%\[ \ket{0} - \ket{1} \] and
%\[\ket{0} + \ket{1}. \]
%The eigenvalues of $U_{f(x)} on
%$\ket{0} + \ket{1}$ are trivial,
%but those of
%$ \ket{0} - \ket{1} $ are equal to $(-1)^{f(x)}$.

Quantum algorithms for approximate counting require roughly only square root
of the number of calls a classical algorithm would require.

\subsubsection{Quantum searching}
While estimating the number of solutions to $f(x) = 1$
is a special case of quantum phase estimation,
the algorithm for searching for these solutions can
be viewed as a clever use of the phase kick-back technique
to induce a desired relative phase between two eigenvectors of $G$.
The state $\ket{X_1}$ is a superposition of solutions to $f(x)=1$, so
it is itself a solution which it is possible for us to construct.

We note that
\begin{equation}
\ket{X_1} = \ket{\psi_+} + \ket{\psi_-}
\end{equation}
and our starting state for quantum searching is
\begin{equation}
 A \ket{0}
= e^{-2 \pi i \theta_j} \ket{\psi_+} + e^{2 \pi i \theta_j} \ket{\psi_-}.
\end{equation}
Each iteration of $G$ kicks back a phase of $e^{2 \pi i \omega_j}$ in front
of $\ket{\psi_+}$
and $e^{-2 \pi i \omega_j}$ in front of
$\ket{\psi_-}$.
So $k$ iterations of $G$ produces the state
\begin{equation}
 A \ket{0}
= \frac{1}{\sqrt{2}}(e^{2 \pi i (k \omega_j - \theta_j )})
\ket{\psi_+} + e^{-2 \pi i (k \omega_j - \theta_j )} \ket{\psi_-}.
\end{equation}
Since we seek
\[ \ket{X_1} = \frac{1}{\sqrt{2}}(\ket{\psi_+} + \ket{\psi_-})  \]
we want to choose the number of iterations $k$ so that
\begin{equation}
k \omega_j
- \theta_j
\end{equation}
is as close to an integer as possible.
When $j$ is small, this means selecting the number of iterations close to
\begin{equation}
\frac{\pi}{4} \sqrt{N/j}.
\end{equation}

Note that any classical algorithm would require $N/j$
evaluations of $f$ before finding a solution to
$f(x) = 1$ with high probability.

\section{Concluding remarks}

Multi-particle interferometers can be viewed as quantum computers and
any quantum algorithm follows the typical structure of a multi-particle
interferometry sequence of operations.
This approach seems to provide
an additional insight into the nature of quantum computation and, we
believe, will help to unify all quantum algorithms and relate them
to different instances of quantum phase estimation.

\section{Acknowledgements}

This work was supported in part by the European TMR Research Network
ERP-4061PL95-1412, Hewlett-Packard and Elsag-Bailey, The Royal
Society, CESG and the Rhodes Trust.
R.C. is partially supported by Canada's NSERC.


\begin{thebibliography}{99}

%
\bibitem{Feynman} R. Feynman: Simulating physics with computers. Int.
  J. Theor. Phys. {\bf 21}, 1982, pp. 467-488.
%
\bibitem{CEMM} R. Cleve, A. Ekert, C. Macchiavello, and M. Mosca:
Quantum Algorithms Revisited, Proc.~R.~Soc.  Lond.  A {\bf
  454}, 1998,
pp. 339--354. See also LANL preprint/quant-ph/9708016.
%
\bibitem {Deutsch85}D. Deutsch: Quantum-theory, the Church-Turing
  principle and the universal quantum computer. Proc.~R.~Soc.  Lond.
  ~A {\bf 400},1985, pp. 97-117.
%
\bibitem{Jones} J. Jones and M. Mosca: Implementation of a quantum
  algorithm on a nuclear-magnetic resonance
  quantum computer. J. Chem. Phys. {\bf 109}, pp. 1648-1653. See also LANL preprint
quant-ph/9801027.
%
\bibitem{Chuang} I. Chuang, L. Vandersypen, X. Zhou, D. Leung and S.
  Lloyd: Experimental realisation of a quantum algorithm. Nature, {\bf
    393}, 1998, pp. 143-146. See also LANL preprint quant-ph/9801037.
%
%
\bibitem{Coppersmith} D. Coppersmith: An Approximate Fourier
Transform Useful in Quantum Factoring, IBM Research Report No.  RC19642,
1994.
%
\bibitem{BEST} A. Barenco, A. Ekert, K. Suominen and P. T\"orma:
Approximate quantum Fourier-transform and decoherence. Phys. Rev. A
{\bf 54}, 1996, pp. 139-146.  See also LANL preprint
quant-ph/9601018.
%
\bibitem{DDEMM} W. van Dam, G. D'Ariano, A. Ekert, C. Macchiavello
and M. Mosca: Estimating Phase Rotations on a Quantum Computer, preprint.
%
\bibitem{Shor}
P.Shor: Algorithms for quantum computation: Discrete logarithms and
factoring. Proc. 35th Annual Symposium on Foundations of Computer
Science, 1994, pp. 124--134.  See also LANL preprint
quant-ph/9508027.
%
\bibitem{EJ}A. Ekert and R. Jozsa: Quantum computation and Shor's
  factoring algorithm, Rev.  Mod.  Phys.  {\bf 68}, 733,
1996, pp. 733-753.
%
\bibitem{MOV} A. Menezes, P. van Oorschot, and S. Vanstone: Handbook of
Applied Cryptography, CRC Press,
London, 1996.
%
\bibitem{Simon}
D. Simon: On the Power of Quantum Computation. Proc. 35th Annual
Symposium on Foundations of Computer Science, 1994, pp. 116-123.
%
\bibitem{Kitaev}
A. Kitaev: Quantum measurements and the Abelian stabiliser problem.
LANL preprint quant-ph/9511026, 1995.
%
\bibitem{Grigoriev}
D. Grigoriev,:
Testing the shift-equivalence of polynomials by
deterministic, probabilistic and quantum machines.
Theoretical Computer Science, {\bf 180}, 1997, pp. 217-228.
%
\bibitem{BL}
D. Boneh, and R. Lipton:
Quantum cryptanalysis of hidden linear functions (Extended abstract).
Lecture Notes on Computer Science, {\bf 963}, 1995, pp.424-437.
%
\bibitem{Mosca.Ekert}
M. Mosca and A. Ekert: Hidden subgroups and estimation of
eigenvalues on a quantum computer. To appear in the Proc. of the
1st International NASA Conference on Quantum Computing and Quantum
Information Processing, Lecture Notes on Computer Science, 1998.
%
\bibitem{Hoyer}
P. H\o yer:
Conjugated Operators in Quantum Algorithms.
preprint, 1997.
%
\bibitem{Grover} L. Grover: A fast quantum mechanical
algorithm for database search, Proc. 28 Annual ACM Symposium on the
    Theory of Computing, ACM Press New York, 1996, pp. 212-219.
    Journal version, ``Quantum Mechanics helps in searching
for a needle in a haystack'', appeared in {\em Physical Review
Letters}, {\bf 79} (1997) 325-328.
    See also LANL preprint quant-ph/9706033.
%
\bibitem{GCK} N. Gershenfeld, I. Chuang and M. Kubinec: Experimental
  implementation of fast quantum searching. Phys. Rev. Lett., {\bf
    80}, 1998, pp. 3408-3411.
%
\bibitem{JHM} J. Jones, R. Hansen and M. Mosca: Implementation of a
  quantum search algorithm on a quantum computer. Nature, {\bf 393},
  1998, pp. 344-346. See also LANL preprint quant-ph/9805069.
%
\bibitem{JM} J. Jones and M. Mosca: Approximate quantum computing
on an NMR ensemble quantum computer. Submitted.  See LANL preprint
quant-ph/quant-ph/9808056.
%
\bibitem{BH} G. Brassard  and P. H{\o}yer: An exact quantum
polynomial-time algorithm for Simon's problem. Proceedings of the
Fifth Israeli Symposium on Theory of Computing and Systems, IEEE
Computer Society Press, 1997, pp.12-23.  See also LANL preprint
quant-ph/9704027.
%
\bibitem{Grover97} L. Grover: A framework for fast
quantum mechanical algorithms. Proc. 30th Annual ACM Symposium on
the
    Theory of Computing, 1998.
See also LANL preprint quant-ph/9711043.
%
\bibitem{BHT} G. Brassard, P. H{\o}yer and A. Tapp:  Quantum Counting,
Proc. 25th International Colloquium on Automata, Languages and
Programming, Lecture Notes on Computer Science, {\bf 1443}, pp.
820-831, 1998. See also LANL preprint quant-ph/9805082.
%

\bibitem{BBHT} M. Boyer, G. Brassard, P. H{\o}yer and A. Tapp: Tight
  bounds on quantum searching, Proceedings of the Fourth Workshop on
  Physics and Computation, 1996, pp. 36-43.
  Forschritte Der Physik, Special issue on quantum computing and quantum cryptography,
  {\bf 4}, pp. 493-505, 1998.  See also LANL preprint quant-ph/9605034.
%
\bibitem{Mosca.Brno}
M. Mosca: Quantum Searching and Counting by Eigenvector Analysis.
Proceedings of Randomized Algorithms, satellite workshop of MFCS
'98. Available at
www.eccc.uni-trier.de/eccc-local/ECCC-LectureNotes/randalg/.
%
\end{thebibliography}
\end{document}